\documentclass{ws-procs9x6}

\begin{document}

\title{ ON THE GRAVITATIONALLY INDUCED SCHWINGER MECHANISM}

\author{GUGLIELMO FUCCI}

\address{Department of Mathematics, Baylor University,\\
Waco, TX 76798, USA\\
E-mail: Guglielmo\textunderscore Fucci@Baylor.edu
}

\author{IVAN G AVRAMIDI}

\address{Department of Mathematics,\\
New Mexico Institute of Mining and Technology\\
Socorro, NM 87801, USA\\
E-mail: iavramid@nmt.edu}

\begin{abstract}
In this paper we will present very recent results obtained in the ambit
of quantum electrodynamics in curved spacetime. We utilize a newly developed non-perturbative
heat kernel asymptotic expansion on homogeneous Abelian bundles over Riemannian manifolds in
order to compute the one-loop effective action for scalar and spinor fields in curved spacetime
under the influence of a strong covariantly constant electromagnetic field. In this framework
we derived, in particular, the gravitational corrections, up to linear terms in Riemannian
curvature, to Schwinger's result for the creation of particles in a strong electric field.
\end{abstract}


\bodymatter

\section{Introduction}\label{sec1}

It is generally recognized that the effective action is a tool of fundamental importance
in quantum field theory and quantum gravity \cite{schwinger51,dewitt65,dewitt03}. In fact,
its knowledge allows one to compute the full propagator and the full vertex functions
of the quantum theory under consideration and, in turn, the $S$-matrix \cite{dewitt65}.
Amongst other methods, the effective action can be computed by utilizing the heat kernel approach
which was first developed by Schwinger  \cite{schwinger51,schwinger54}, and
later generalized to include curved spacetime by DeWitt \cite{dewitt67a,dewitt67b,dewitt75}.
In particular Schwinger computed the effective action for constant electromagnetic fields
in Minkowski spacetime \cite{schwinger51}. He noticed that in presence of an electric field
the effective action acquires an imaginary part interpreted as creation of pairs induced by the
electric field. This effect has been since then known as Schwinger mechanism.
From a formal point of view the presence of an imaginary part of the effective
action can be understood as follows: the effective action is given in terms of a particular
integral, properly regularized, over $t$ of the heat kernel diagonal \cite{dewitt65,dewitt03}.
In the presence of an electric field the heat kernel diagonal becomes a meromorphic function
with isolated single poles on the real axis. These poles are, then, avoided by deforming
the contour of integration which leads to an imaginary part given by the sum of the residues
of all the poles \cite{schwinger51}.

In this paper we will utilize a newly developed non-perturbative heat kernel asymptotic expansion for
homogeneous Abelian bundles in order to obtain the gravitational correction (up to the first order in the
Riemannian curvature) to the Schwinger mechanism for a covariantly constant electric field.

\section{Non-perturbative Heat Kernel Asymptotic Expansion}\label{sec2}

Let $(M,g)$ be a $n$-dimensional Riemannian manifold
without boundary and $\mathcal{S}$ be a complex vector bundle over $M$ realizing
a representation of the group $G\otimes U(1)$, where $G$ is a compact semisimple Lie group. Let
$\nabla$ be the total connection on the vector bundle $\mathcal{S}$,
${\cal R}_{\mu\nu}$ be the curvature of the $G$-connection
and $F_{\mu\nu}$ be the curvature of the $U(1)$-connection (which will be
called the electromagnetic field).

Let $U(t;x,x')$ be the heat kernel of the Laplacian
$L=-g^{\mu\nu}\nabla_\mu\nabla_\nu$ and $\Theta(t)=U(t;x,x)$ be the heat kernel diagonal. As $t\to 0$ the heat kernel diagonal has a well known asymptotic
expansion
\begin{equation}
\Theta(t)\sim(4\pi t)^{-n/2}\sum_{k=0}^\infty t^k a_k\,,
\label{4}
\end{equation}
where $a_k$ are the well known local heat kernel coefficients, which are polynomials
in the curvatures (both $R$ and $F$) and their derivatives.

In Ref.~\refcite{avramidi08e} we have considered the case in which
\begin{equation}
R\ll F\;,\quad \nabla\nabla R\ll F^{2}\;,\quad \nabla F=0\;.
\label{5}
\end{equation}
In this situation the electromagnetic field cannot be treated as a perturbation
and instead of (\ref{4}) we obtained
a {\it new, non-perturbative}
expansion of the heat kernel diagonal
\cite{avramidi08e}
\begin{eqnarray}
\Theta(t)&\sim&(4\pi t)^{-n/2}\sum_{k=0}^{\infty}t^{k} b_k(t)\;.
\label{6}
\end{eqnarray}
The coefficients
$b_k(t)$
are polynomials in the Riemann curvature and its derivatives with the coefficients that are
some universal dimensionless tensor-valued
analytic functions that depend on $F$ only in the dimensionless combination
$tF$.
These new non-perturbative coefficients of the heat kernel asymptotic expansion
have been computed in
Ref.~\refcite{avramidi08e} by utilizing a promising algebraic approach developed in
Refs.~\refcite{avramidi93,avramidi95,avramidi95a,avramidi96}. The form of the
coefficients of the asymptotic expansion for the heat kernel diagonal can be expressed
as follows
\begin{eqnarray}
b_{2}(t)&=&\frac{1}{6}R+\Psi^{\mu\nu}(t)R_{\mu\nu}\;,\qquad b_{3}(t)=0\;,\\
b_{4}(t)&=&-\frac{1}{72}R^{2}+\frac{1}{6}R\;b_{2}(t)+\Phi_{1}^{\mu\nu\alpha\beta\gamma\delta\tau\sigma}(t)\;R_{\mu\nu\alpha\beta}R_{\gamma\delta\tau\sigma}\nonumber\\
&+&\Phi_{2}^{\mu\nu\alpha\beta\gamma\delta}(t)\nabla_{\mu}\nabla_{\nu}R_{\alpha\beta\gamma\delta}\;,
\label{7}
\end{eqnarray}
where $\Psi(t)$, $\Phi_{1}(t)$ and $\Phi_{2}(t)$ are quite involved analytic tensorial functions of $F$
which have been explicitly obtained in  Ref.~\refcite{avramidi08e}.

\section{One-Loop Effective Action and its Imaginary Part}\label{sec3}

The heat kernel asymptotic expansion described in the previous section has been used
in Ref.~\refcite{avramidi09a} in order to evaluate the one-loop effective action, up to linear orders in the Riemannian curvature,
for scalar and spinor fields under the influence of a strong covariantly constant electromagnetic field
in curved spacetime. In the framework of $\zeta$-function regularization, one can write the one-loop
effective Lagrangian in terms of the heat kernel diagonal as follows \cite{hawking77}
\begin{equation}
{\cal L}=-\sigma
\int\limits_{\epsilon/\mu^{2}}^{\infty}\frac{dt}{t}e^{-tm^{2}}\Theta(t)\;,
\label{8}
\end{equation}
where $\sigma=+1$ for bosons and $\sigma=-1$ for fermions,
$m$ is the mass of the field and $\epsilon$ and $\mu$ are, respectively,
the regularization and renormalization parameters.

As pointed out before, in presence of an electric field the heat kernel diagonal
$\Theta(t)$
yields poles on the positive real axis which contribute to the imaginary part of the effective
Lagrangian through their residues as \cite{avramidi09a}
\begin{equation}
{\rm Im}\;{\cal L}
=-\sigma \pi\sum_{k=1}^\infty
\textrm{Res}\left\{t^{-1} e^{-t m^2} \Theta(t); t_k\right\}\,.
\label{10}
\end{equation}
By utilizing the spectral decomposition of the electromagnetic $2$-form $F$,
it was shown in Ref.~\refcite{avramidi09a} that the imaginary part of the effective
Lagrangian for scalar and spinor fields under the influence of solely an electric field ${\cal E}$
has the form
\begin{eqnarray}
\textrm{Im}\;\mathcal{L}
&=&\pi(4\pi)^{-\frac{n}{2}}E^{\frac{n}{2}}G_{0}(y)
+\pi(4\pi)^{-\frac{n}{2}}E^{\frac{n}{2}-1}
\big[G_{1}(y)R+G_{2}(y)\Pi_{1}^{\mu\nu}R_{\mu\nu}\nonumber\\
&+&G_{4}(y)\Pi_1^{\mu\nu}\Pi_1^{\alpha\beta}R_{\mu\alpha\nu\beta}+G_{5}(y)E_1^{\mu\alpha}E_1^{\nu\beta}R_{\mu\alpha\nu\beta}\big]\;,
\label{11}
\end{eqnarray}
where $y=m^{2}/{\cal E}$, $E_{k}$ are antisymmetric matrices satisfying \cite{avramidi09a}
\begin{equation}
E_k{}_{\mu\nu}=-E_k{}_{\nu\mu}\;,\;\;E^k_{\mu[\nu}E^k_{\alpha\beta]}=0\;,\;\;E_k E_{m}=0\;,\;\;\textrm{(for}\, k\neq m)
\end{equation}
and the projectors $\Pi_{k}$ are defined by \cite{avramidi09a}
\begin{equation}
\Pi_k=-E_k^2\;.\;\;
\end{equation}
One can easily recognize that $G_{0}$ in the expression (\ref{11}) represents the
original term computed by Schwinger, \cite{schwinger51} while the additional functions
proportional to the Riemannian curvature have been explicitly evaluated in arbitrary dimensions
in Ref.~\refcite{avramidi09a} and represent the {\it new} gravitational contribution to the Schwinger mechanism.

In the physically relevant case of a four-dimensional spacetime one obtains
for scalar fields \cite{avramidi09a}
\begin{equation}
G_{0}^{\rm scalar}(y)
=-\frac{1}{\pi^{2}}\textrm{Li}_{2}(-e^{-\pi
y})\;,\;\;G_{1}^{\rm scalar}(y)=
-\left(\frac{1}{6}-\xi\right)
\frac{1}{\pi}\ln(1+e^{-\pi y})\;,
\label{12}
\end{equation}
\begin{eqnarray}
G_{2}^{\rm scalar}(y)&=&\frac{1}{48\pi^{3}}
\bigg\{\frac{2\pi^{3}y e^{-\pi y}}{1+e^{-\pi y}}
+8\pi^{2}\ln(1+e^{-\pi y})
\nonumber\\
&+&18\pi y\textrm{Li}_{2}(-e^{-\pi y})
+54\textrm{Li}_{3}(-e^{-\pi y})\bigg\}\;,
\end{eqnarray}
\begin{eqnarray}
G_{4}^{\rm scalar}(y)&=&
\frac{1}{384\pi^{3}}
\bigg\{\frac{16\pi^{3}y e^{-\pi y}}{1+e^{-\pi y}}
+4\pi^{2}(17-3y^{2})\ln(1+e^{-\pi y})
\nonumber\\
&+&192\pi y\textrm{Li}_{2}(-e^{-\pi y})
+504\textrm{Li}_{3}(-e^{-\pi y})\bigg\}\;,
\end{eqnarray}
\begin{eqnarray}
G_{5}^{\rm scalar}(0,y)&=&
-\frac{1}{256\pi^{3}}
\big\{4\pi^{2}(1-3y^{2})\ln(1+e^{-\pi y})
+48\pi y\textrm{Li}_{2}(-e^{-\pi y})
\nonumber\\
&+&72\textrm{Li}_{3}(-e^{-\pi y})\big\}\;,
\label{13}
\end{eqnarray}
where $\xi$ represents the coupling constant,
while for spinor fields one has \cite{avramidi09a}
\begin{equation}
G_{0}^{\rm spinor}(y)=
\frac{4}{\pi^{2}}\textrm{Li}_{2}(e^{-\pi
y})\;,\;\;G_{1}^{\rm spinor}(y)=
\frac{1}{3\pi}\ln(1-e^{-\pi y})\;,
\end{equation}
\begin{eqnarray}
G_{2}^{\rm spinor}(y)
&=&-\frac{1}{12\pi^{3}}\bigg\{\frac{2\pi^{3}y
e^{-\pi y}}{1-e^{-\pi y}}
-8\pi^{2}\ln(1-e^{-\pi y})
\nonumber\\
&-&18\pi y\textrm{Li}_{2}(e^{-\pi y})
-54\textrm{Li}_{3}(e^{-\pi y})\bigg\}\;,
\end{eqnarray}
\begin{eqnarray}
G_{4}^{\rm spinor}(y)&=&
-\frac{1}{96\pi^{3}}\bigg\{\frac{16\pi^{3}y e^{-\pi y}}{1-e^{-\pi y}}
-4\pi^{2}(20-3y^{2})\ln(1-e^{-\pi y})
\nonumber\\
&-&192\pi y\textrm{Li}_{2}(e^{-\pi y})
-504\textrm{Li}_{3}(e^{-\pi y})\bigg\}\;,
\end{eqnarray}
\begin{eqnarray}
G_{5}^{\rm spinor}(y)&=&\frac{3}{16\pi^{3}}
\big\{\pi^{2}(4+y^{2})\ln(1-e^{-\pi y})
-4\pi y\textrm{Li}_{2}(e^{-\pi y})
\nonumber\\
&-&6\textrm{Li}_{3}(e^{-\pi y})\big\}\;,
\end{eqnarray}
where $\textrm{Li}_{j}(z)$ in the above formulas represents the polylogarithmic function.
It is interesting, at this point, to analyze the limit as $y$ approaches zero, namely
the situation in which ${\cal E}\gg m^{2}$. For a scalar field one obtains \cite{avramidi09a}, by taking the limit $y\to 0$
of the expressions  (\ref{12})-(\ref{13}),
\begin{equation}
G_{0}^{\rm scalar}=\frac{1}{12}\;,\;\;G_{1}^{\rm scalar}
=-\left(\frac{1}{6}-\xi\right)\frac{1}{\pi}\ln 2\;,
\end{equation}
\begin{equation}
G_{2}^{\rm scalar}
=\frac{1}{6\pi}\ln 2 -\frac{27}{32 \pi^{3}}\zeta(3)\;,\;\;G_{4}^{\rm scalar}
=\frac{17}{96\pi}\ln 2-\frac{63}{64\pi^{3}}\zeta(3)\;,
\end{equation}
\begin{equation}
G_{5}^{\rm scalar}
=-\frac{1}{64\pi}\ln 2+\frac{27}{128\pi^{3}}\zeta(3)\;,
\end{equation}
where $\zeta(s)$ is the Riemann zeta function.
$G_{i}^{\rm spinor}(y)$, for $i\neq 0$, in
four dimensions represent a special case since there is an infrared
divergence as $m\rightarrow 0$ (or $y\rightarrow 0$). This means that there
is no well-defined value for the small mass limit. Instead,
one finds a logarithmic divergence, $\log (\pi y)$, as follows \cite{avramidi09a}
\begin{equation}
G_{0}^{\rm spinor}=\frac{2}{3}\;,\;\;G_{1}^{\rm spinor}(y)
=\frac{1}{3\pi}\log(\pi y)+O(y)\;,
\end{equation}
\begin{equation}
G_{2}^{\rm spinor}(y)
=\frac{2}{3\pi}\log(\pi y)
-\frac{1}{6\pi}
+\frac{9}{2\pi^{3}}\zeta(3)+O(y)\;,
\end{equation}
\begin{equation}
G_{4}^{\rm spinor}(y)
=\frac{5}{6\pi}\log(\pi y)
-\frac{1}{6\pi}
+\frac{21}{4\pi^{3}}\zeta(3)+O(y)\;,
\end{equation}
\begin{equation}
G_{5}^{\rm spinor}(y)
=\frac{3}{4\pi}\log(\pi y)
+\frac{9}{8\pi^{3}}\zeta(3)+O(y)\;.
\end{equation}
From the last formulas, one can clearly see that there are {\it new infrared divergences} in the
imaginary part of the effective action for spinor fields in four dimensions when ${\cal E}\gg m^{2}$.
This means, physically, that the creation of massless spinor particles
(or massive particles in supercritical electric field) is magnified substantially by the
presence of the gravitational field \cite{avramidi09a}. We would like to mention, here, that
similar infrared divergences appear in the real part of the effective Lagrangian for massless
spinor fields under the influence of a pure magnetic field. This means that the vacuum energy
of charged spinors with small mass (or equivalently massive charged spinors for
which $m^{2}\ll {\cal B}$) dramatically increases due to the presence of the gravitational field \cite{fucci09}.

\section{Conclusions}

In this paper we have briefly described interesting new developments in the
ambit of quantum electrodynamics in arbitrarily curved spacetimes. In particular,
by using a new non-perturbative heat kernel asymptotic expansion obtained in  Ref.~\refcite{avramidi08e},
we have found, in Ref.~\refcite{avramidi09a}, the gravitational corrections to the creation of pairs in a strong electric field.
The Schwinger mechanism has, in fact, lately gained importance in light of new experiments
proposed in order to observe the creation of electron-positron pairs from vacuum \cite{dunne08,dunne09,hebe09}.
It would certainly be of interest to specialize the gravitational corrections to the Schwinger mechanism
to gravitational backgrounds which are relevant in astrophysical settings. This would be important in order
to understand in which situations the gravitationally induced Schwinger mechanism might be observable
\footnote{We would like to thank Christian Schubert for suggesting this problem.}.
This effect could also have important
consequences for theories with spontaneous breakdown of symmetry when the mass of
charged particles is generated by a Higgs field. These theories would exhibit an
enhancement of created particles (in the massless limit an infinite
amount) at the phase transition point when the symmetry is restored and the
massive charged particles become massless.
This seems to be an interesting {\it new physical
effect} that, for the reasons mentioned before, deserves further investigation.

\section*{Acknowledgments}

GF would like to thank the organizers of QFEXT09 for such an excellent conference
and the Department of Mathematics at Baylor University for the financial support to attend the conference.

\end{document}